\documentclass[12pt,leqno]{article}

\title{Estimate of blow-up and relaxation time for
self-gravitating Brownian particles and bacterial populations}

\usepackage{amsthm,amsmath,amssymb,a4wide,psfig,epsf}
\def\mb#1{\setbox0=\hbox{$#1$}\kern-.025em\copy0\kern-\wd0
\kern-0.05em\copy0\kern-\wd0\kern-.025em\raise.0233em\box0}

\begin{document}

\author{P.-H. Chavanis and C. Sire}
\maketitle
\begin{center}
Laboratoire de Physique Th\'eorique (UMR 5152 du CNRS), Universit\'e
Paul Sabatier,\\ 118, route de Narbonne, 31062 Toulouse Cedex 4, France\\
E-mail: {\it chavanis{@}irsamc.ups-tlse.fr, clement.sire@irsamc.ups-tlse.fr }

\vspace{0.5cm}
\end{center}

\begin{abstract}

We determine an asymptotic expression of the blow-up time $t_{coll}$
for self-gravitating Brownian particles or bacterial populations
(chemotaxis) close to the critical point. We show that
$t_{coll}=t_{*}(\eta-\eta_{c})^{-1/2}$ with $t_{*}=0.91767702...$,
where $\eta$ represents the inverse temperature (for Brownian
particles) or the mass (for bacterial colonies), and $\eta_{c}$ is the
critical value of $\eta$ above which the system blows up. This result
is in perfect agreement with the numerical solution of the
Smoluchowski-Poisson system. We also determine the asymptotic
expression of the relaxation time close but above the critical
temperature and derive a large time asymptotic expansion for the
density profile exactly at the critical point.

\end{abstract}

\section{Introduction}
\label{sec_introduction}

Recently, several papers have focused on the blow-up properties of a
cluster of self-attracting particles
\cite{wolansky,biler1,csr,herrero,dolbeault,rosier,
crs,sc,biler2,anomalous,sc2,guerra}. This concerns in particular the
``isothermal collapse'' \cite{aa} of self-gravitating Brownian
particles and the ``chemotactic aggregation'' \cite{ks} of bacterial
populations in biology. These systems have a similar mathematical
structure and they are described, in a first approximation, by the
Smoluchowski-Poisson system \cite{crrs}. The Smoluchowski equation is
a particular Fokker-Planck equation in physical space involving a
diffusion due to Brownian motion and a drift. In the standard Brownian
theory developed by Einstein, the drift is due to an external
potential \cite{risken}. Alternatively, we can consider the more
complicated situation in which the potential is produced by the
particles themselves. If we assume a Newtonian type of interaction, we
have to couple the Smoluchowski equation to the Poisson equation.

It can be shown that the Smoluchowski-Poisson system decreases a
Lyapunov functional $F[\rho]$ which can be interpreted as a Boltzmann
free energy \cite{gtpre}. Furthermore, the type of evolution depends
on the value of a dimensionless parameter $\eta$. For
$\eta\leq\eta_{c}=2.51755132...$, the Smoluchowski-Poisson system
evolves to an equilibrium state which minimizes the free energy at
fixed mass. It corresponds to an isothermal distribution of particles
similar to isothermal stars and isothermal stellar systems
\cite{bt}. On the contrary, for $\eta>\eta_{c}$, there is no possible
equilibrium state and the system blows up. This is the case for
self-gravitating Brownian particles below a critical temperature
$T_{c}$ and for bacterial populations above a critical mass
$M_{c}$. It is found that the collapse is self-similar and that it
develops a finite time singularity, {\it i.e.} the central density
becomes infinite in a finite time $t_{coll}$ \cite{crs,sc}. Then, a
Dirac peak is formed in the post-collapse regime \cite{sc2}.

It is clear that the collapse time $t_{coll}$ depends on the
distance to the critical point $\eta_{c}$ and that it should diverge as
$\eta\to \eta_{c}$.  More precisely, by using heuristic
arguments, it is argued in \cite{crs} that $t_{coll}\sim t_{*}
(\eta_{c}-\eta)^{-1/2}$. This scaling is consistent with numerical
simulations of the Smoluchowski-Poisson system. However, the
approach of \cite{crs} is qualitative and does not provide the
numerical value of $t_{*}$.

After some introductory material presented in Section 2, we
develop a systematic procedure in Section 3, inspired from \cite{kavallaris},
which confirms the scaling law $t_{coll}\sim t_*
(\eta_{c}-\eta)^{-1/2}$ and leads to the explicit value of
$t_{*}$. In section 4, we study the relaxation time $\tau$ to
equilibrium below $\eta_c$ and show that it diverges like
$\tau\sim c_\eta (\eta-\eta_{c})^{-1/2}$, where $c_\eta$ is given
explicitly. In Section 5, we derive a systematic large
time expansion of the density profile exactly at $\eta_c$. In
particular, we show that the central density approaches its
equilibrium value according to $\rho_0-\rho(r=0,t)\sim
\frac{c_\rho}{t}$, where $c_\rho$ is a universal constant which is
explicitly computed. Finally, in Section 6, we determine the pulsation
period of bounded isothermal spheres described by the Euler-Jeans
equations close to the critical point.

\section{Self-gravitating Brownian particles and bacterial populations}

\subsection{The model equations}

In the overdamped regime, a system of self-gravitating Brownian
particles is described by the $N$ coupled stochastic equations
\begin{eqnarray}
{d{\bf r}_i\over dt}=-\mu\nabla_i U({\bf r}_1,...,{\bf r}_N)+\sqrt{2D}\ {\bf
R}_i(t),
 \label{brown3}
\end{eqnarray}
where $\mu=1/\xi$ is the mobility ($\xi$ is the friction coefficient),
$D$ is the diffusion coefficient and ${\bf R}_{i}(t)$ is a white noise
satisfying $\langle {\bf R}_{i}(t)\rangle={\bf 0}$ and $\langle
{R}_{a,i}(t){R}_{b,j}(t')\rangle=\delta_{ij}\delta_{ab}\delta(t-t')$,
where $a,b=1,2,3$ refer to the coordinates of space and $i,j=1,...,N$
to the particles. We define the temperature $T=1/\beta$ through the
Einstein relation $\mu=D\beta$. The particles interact via the
potential $U({\bf r}_{1},...,{\bf r}_{N})=\sum_{i<j}u({\bf r}_{i}-{\bf
r}_{j})$. In this paper, $u({\bf r}_{i}-{\bf r}_{j})=-G/|{\bf
r}_{i}-{\bf r}_{j}|$ is the Newtonian binary potential in $d=3$
dimensions. Starting from the $N$-body Fokker-Planck equation and
using a mean-field approximation
\cite{martzel,bedlewo}, we can derive the nonlocal Smoluchowski
equation
\begin{eqnarray}
\label{brown5} {\partial\rho\over\partial t}=\nabla \biggl \lbrack
{1\over\xi}(T\nabla\rho+\rho\nabla\Phi)\biggr\rbrack,
\end{eqnarray}
where the potential $\Phi({\bf r},t)$ is produced by the density
$\rho({\bf r},t)$ according to the Poisson equation
\begin{eqnarray}
\Delta\Phi=4\pi G\rho.
 \label{po}
\end{eqnarray}
The self-gravitating Brownian gas has a rigorous canonical
thermodynamic structure \cite{crs}. It can be seen therefore as the
canonical counterpart of the usual $N$-stars problem governed by
Newton's equations and possessing a microcanonical structure.

On the other hand, in a biological context, a model of chemotactic
aggregation exhibiting blow-up phenomena is provided by the system of
equations
\begin{equation}
\label{ge5}
{\partial\rho\over\partial t}=D\Delta\rho-\chi\nabla (\rho\nabla c),
\end{equation}
\begin{equation}
\label{ge6}\Delta c=-\lambda\rho,
\end{equation}
where $\rho$ is the concentration of the biological population
(amoebae), $c$ is the concentration of the substance secreted
(acrasin) and $\chi$ measures the strength of the chemotactic
drift. Equations (\ref{ge5}) and (\ref{ge6}) can be viewed as a
simplification of the Keller-Segel model \cite{ks}. Clearly, the two
models are isomorphic provided that we make the identification
$\Phi\leftrightarrow -{4\pi G\over\lambda}c$, $T\leftrightarrow {4\pi
GD\over\lambda\chi}$ and $\xi \leftrightarrow {4\pi G\over
\lambda\chi}$. Introducing the mass $M=\int\rho \,d^{3}{\bf r}$ of the
system and the radius $R$ of the domain, we can show that the
problem depends on the single dimensionless parameter $\eta=\beta
GM/R$ \cite{crs}.  Therefore, a large value of $\eta$ corresponds to a small
temperature $T$ or a large mass $M$.

\subsection{The Smoluchowski-Poisson system}

In the following, we shall use the gravitational terminology but we
stress that our results are equally valid for the biological problem
due to the above analogy. By rescaling the physical parameters
adequately, it is possible to take $M=R=G=\xi=1$ without restriction.
Then, the equations of the problem are
\begin{eqnarray}
{\partial\rho\over\partial t}&=&\nabla (T\nabla\rho+\rho\nabla\Phi),
\label{dim1}\\
\Delta\Phi&=&4\pi\rho,
\label{dim2}
\end{eqnarray}
with proper boundary conditions in order to impose a vanishing
particle flux on the surface of the confining sphere. If we restrict
ourselves to spherically symmetric solutions, the boundary conditions
are
\begin{equation}
{\partial\Phi\over\partial r}(0,t)=0, \qquad \Phi(1,t)=-1,
\qquad T{\partial \rho\over\partial r}(1,t)+\rho(1,t)=0.
\label{dim4}
\end{equation}
With this rescaling, the equations only depend on the temperature $T$,
or equivalently on the parameter $\eta=1/T$.  As mentioned previously, a
small temperature is equivalent (through a rescaling) to a large mass.

Integrating Eq.~(\ref{dim2}) once, we can rewrite the
Smoluchowski-Poisson system in the form of a single
integrodifferential equation
\begin{equation}
{\partial\rho\over\partial t}={1\over r^{2}}{\partial\over\partial
r}\biggl\lbrace r^{2}\biggl (T{\partial\rho\over\partial
r}+{\rho\over r^{2}}\int_{0}^{r}\rho(r',t)4\pi r^{'2}\,dr'\biggr
)\biggr \rbrace.
\label{dim5a}
\end{equation}
The Smoluchowski-Poisson system is also equivalent to a single
differential equation
\begin{equation}
\frac{\partial M}{\partial t}=T \left(\frac{\partial^2 M}{\partial r^2}
-\frac{2}{r}\frac{\partial M}{\partial r}\right)
+{1\over r^{2}}M\frac{\partial M}{\partial r},
\label{sca}
\end{equation}
for the quantity
\begin{equation}
M(r,t)=\int_{0}^{r}\rho(r',t)4\pi r^{'2}\,dr', \label{dintm}
\end{equation}
which represents the mass contained within the sphere of radius
$r$. The appropriate boundary conditions are
\begin{equation}
 M(0,t)=0,\qquad M(1,t)=1.
\label{dintb}
\end{equation}

\section{Asymptotic estimate of the collapse time }

\subsection{Systematic expansion}

We know that the Smoluchowski-Poisson system blows up for
$T<T_{c}$, where $T_{c}$ is a critical temperature below which
there is no equilibrium state \cite{crs}. We shall place ourselves close to
the critical point and expand the mass profile as
\begin{equation}
M(r,t)=M_{c}(r)+\epsilon M_{1}(r,t)+\epsilon^{2}M_{2}(r,t)+...,\label{dint1}
\end{equation}
where $M_{c}(r)$ is the equilibrium profile at $T_{c}$ and $\epsilon$
is a small parameter defined as
\begin{equation}
\epsilon=\biggl ({T_{c}-T\over T_{c}}\biggr )^{1/2}\ll 1.
\label{dint2}
\end{equation}
We also rescale the time according to
\begin{equation}
t=\tau/\epsilon.
\label{dint3}
\end{equation}
If we substitute the expansion Eq.~(\ref{dint1}) in Eq.~(\ref{sca}) and
equal terms of same order, we get to order $0$:
\begin{equation}
T_{c}\biggl ({\partial^{2}M_{c}\over\partial r^{2}}-{2\over
r}{\partial M_{c}\over\partial r}\biggr )+{M_{c}\over r^{2}}{\partial
M_{c}\over\partial r}=0,
\label{dint4}
\end{equation}
to order $1$:
\begin{equation}
T_{c}\biggl ({\partial^{2}M_{1}\over\partial r^{2}}-{2\over
r}{\partial M_{1}\over\partial r}\biggr )+{1\over r^{2}} {\partial
\over\partial r}(M_{1}M_{c})=0,
\label{dint5}
\end{equation}
and to order $2$:
\begin{equation}
{\partial M_{1}\over\partial\tau}=T_{c}\biggl
({\partial^{2}M_{2}\over\partial r^{2}}-{2\over r}{\partial
M_{2}\over\partial r}-{\partial^{2}M_{c}\over\partial r^{2}}+{2\over
r}{\partial M_{c}\over\partial r}\biggr )+{1\over r^{2}} \biggl
(M_{c}{\partial M_{2}\over\partial r}+M_{1}{\partial
M_{1}\over\partial r}+M_{2}{\partial M_{c}\over\partial r}\biggr ).
\label{dint6}
\end{equation}
The boundary conditions are
\begin{equation}
M_{n\ge 0}(0,\tau)=0, \quad M_{n\ge 0}'(0,\tau)=0,
\label{dint7}
\end{equation}
\begin{equation}
M_{c}(1,\tau)=1, \quad M_{n\ge 1}(1,\tau)=0.
\label{dint8}
\end{equation}

\begin{figure}[htbp]
\vskip1.5cm
\centerline{
\psfig{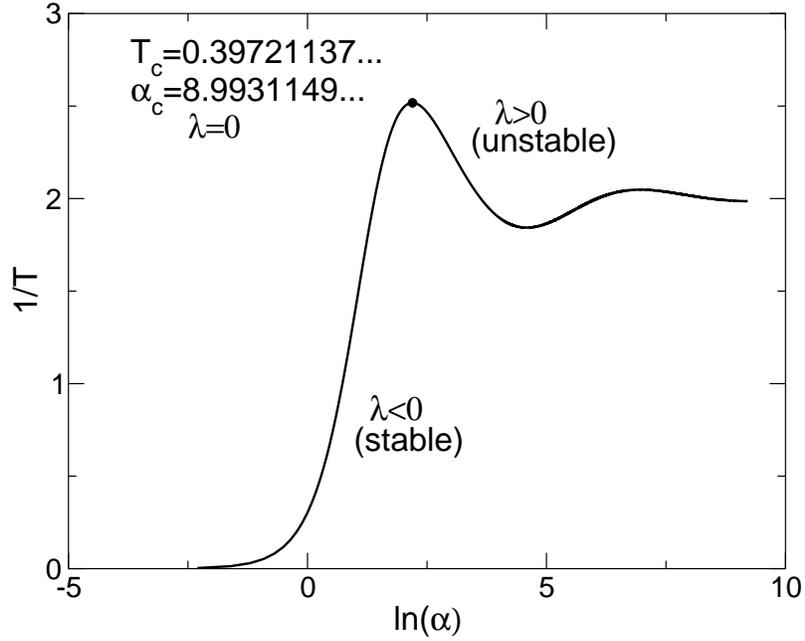}}
\caption{We plot $\eta(\alpha)=T^{-1}(\alpha)$ where $\alpha=(4\pi\rho_{0}/T)^{1/2}$ parameterizes the series of equilibria. For $\alpha>\alpha_c$,
unstable modes with $\lambda>0$ arise, which are solution of
Eq.~(\ref{dint20}). The minimum temperature $T_{c}$ corresponds
precisely at the point of marginal stability $\lambda=0$.}
\label{alphaT}
\end{figure}

\begin{figure}[htbp]
\vskip1.5cm
\centerline{
\psfig{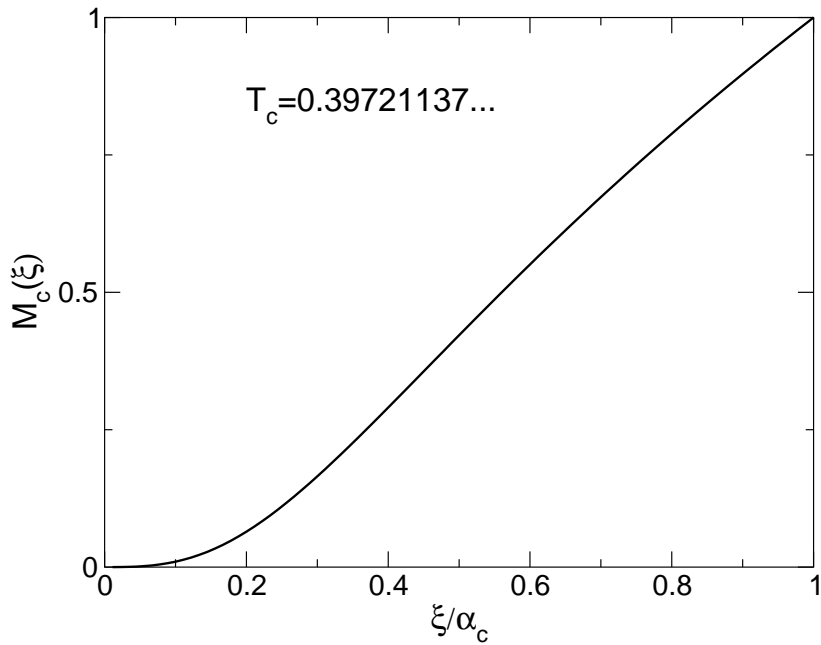}}
\caption{The integrated mass density is plotted at the critical point. }
\label{xiM}
\end{figure}

\subsection{Order $0$: the equilibrium state $M_{c}(\xi)$}

At equilibrium, the condition of hydrostatic balance
$T\nabla\rho+\rho\nabla\Phi=0$ combined with the Gauss theorem
$d\Phi/dr=M(r)/r^{2}$ leads to the relation
\begin{equation}
M(r)=-Tr^{2}{d\ln\rho\over dr}.
\label{dint9}
\end{equation}
Since $M'(r)=4\pi \rho r^{2}$, we obtain the differential equation
\begin{equation}
{1\over r^{2}}{d\over dr}\biggl (r^{2}{T\over\rho}{d\rho\over
dr}\biggr )=-4\pi \rho. \label{dint10}
\end{equation}
Introducing the function $\psi$ through the relation
\begin{equation}
\rho=\rho_{0}e^{-\psi},
\label{dint11}
\end{equation}
where $\rho_{0}$ is the central density, and using the normalized
distance $\xi=\alpha r$ where $\alpha=(4\pi\rho_{0}/T)^{1/2}$, we can
rewrite the relation Eq.~(\ref{dint9}) in the form
\begin{equation}
M(\xi)=T{\xi^{2}\over\alpha}\psi'(\xi).
\label{dint12}
\end{equation}
Furthermore, according to Eq.~(\ref{dint10}), the function $\psi$ is
solution of the Emden equation
\begin{equation}
{1\over\xi^{2}}{d\over d\xi}\biggl (\xi^{2}{d\psi\over d\xi}\biggr )=e^{-\psi},
\label{emden}
\end{equation}
\begin{equation}
\psi(0)=\psi'(0)=0.
\label{dint13}
\end{equation}
The Taylor expansion of the Emden function near the origin is
\begin{equation}
\psi={1\over 6}\xi^{2}-{1\over 120}\xi^{4}+...
\label{zero7}
\end{equation}
Using these results, we can check that the Emden equation (\ref{emden}) is
equivalent to Eq. (\ref{dint4}), as it should. For a given temperature $T\ge T_{c}$, the Emden equation has to be
solved from $\xi=0$ to $\xi=\alpha$ such that
\begin{equation}
\alpha\psi'(\alpha)=\frac{1}{T}.
\label{dint15}
\end{equation}
For the profile we are considering, at the critical temperature
$T_{c}$, we have to stop the integration at $\alpha_{c}$ such that
$T(\alpha_{c})$ is minimum. The condition $[1/T]'(\alpha_{c})=0$
is equivalent to 
\begin{equation}
{\alpha_{c}\ e^{-\psi(\alpha_{c})}\over \psi'(\alpha_{c})}=1.
\label{dint16}
\end{equation}
It is found numerically that
$\alpha_{c}=8.9931149...$ and $T_{c}=0.39721137...$.

\subsection{Order $1$: the function $F(\xi)$}

Since  Eq.~(\ref{dint5}) is linear, we look for solutions of the form
\begin{equation}
M_{1}=A(\tau)F(\xi).
\label{dint17}
\end{equation}
The function $F(\xi)$ satisfies the differential equation
\begin{equation}
{d^{2}F\over d\xi^{2}}-{2\over\xi}{dF\over
d\xi}+{1\over\xi^{2}}{d\over d\xi}(F\xi^{2}\psi')=0,
\label{dint18}
\end{equation}
which is equivalent to
\begin{equation}
{d\over d\xi}\biggl ({e^{\psi}\over \xi^{2}}{dF\over d\xi}\biggr
)+{F\over\xi^{2}}=0,
\label{dint19}
\end{equation}
where we have used the Emden equation Eq.~(\ref{emden}).
Equation (\ref{dint19}) is precisely the equation found at the critical
point $T_{c}$ by analyzing the linear stability of isothermal
spheres \cite{crs}. Indeed, considering a perturbation $\delta M\sim e^{\lambda
t}$ around a stationary solution $M(r)$ of Eq.~(\ref{sca}), we get
\begin{equation}
T\biggl ({\partial^{2}\delta M\over\partial r^{2}}-{2\over
r}{\partial \delta M\over\partial r}\biggr )+{1\over r^{2}} {\partial
\over\partial r}(M\delta M)=\lambda \delta M.
\label{dint20}
\end{equation}
If $\lambda<0$, the stationary solution is stable and if $\lambda>0$,
the stationary solution is unstable. At the critical point $T_{c}$ where
$\lambda=0$ (marginal stability), we recover Eq.~(\ref{dint5}) with
$M_{1}=\delta M$. Equation (\ref{dint19}) has to be solved with the
boundary conditions
\begin{equation}
F(0)=F'(0)=0.
\label{dintF0}
\end{equation}
We find that $F(\xi)$ can be simply expressed in terms of 
$\psi(\xi)$ as \cite{aa}:
\begin{equation}
F(\xi)=\xi^{3}e^{-\psi}-\xi^{2}\psi'.
\label{dint21}
\end{equation}
The condition $F(\alpha_{c})=0$ determines the critical
temperature $T_{c}$. Using Eq. (\ref{dint21}), this condition is equivalent to
\begin{equation}
{\alpha_{c}\ e^{-\psi(\alpha_{c})}\over \psi'(\alpha_{c})}=1.
\label{dintconda}
\end{equation}
Comparing this result with Eq. (\ref{dint16}), we find that the criterion of marginal stability ($\lambda=0$) corresponds to the point $\alpha_{c}$ in the series of equilibria $T=T(\alpha)$ where the temperature is minimum ($[1/T]'(\alpha_{c})=0$) \cite{aa,crs}.

\subsection{Order $2$: the function $A(\tau)$  }
\label{sec_two}

Using the previous results,  Eq.~(\ref{dint6}) can be rewritten
\begin{equation}
F(\xi)\dot A(\tau)=T_{c}\alpha_c^{2}\biggl
\lbrack{\partial^{2}M_{2}\over\partial\xi^{2}}-{2\over\xi}{\partial
M_{2}\over\partial
\xi}+{1\over\xi^{2}}{\partial\over\partial\xi}(M_{2}\xi^{2}\psi')\biggr
\rbrack+\alpha_c
T_{c}^{2}\xi^{2}\psi'e^{-\psi}+{\alpha_c^{3}\over\xi^{2}}A^{2}F{dF\over
d\xi}.
\label{dint22}
\end{equation}
We now multiply Eq.~(\ref{dint22}) by a function $\chi(\xi)$ and integrate
between $0$ and $\alpha_c$. We choose
the function $\chi$ (see below) so as to eliminate the terms where
$M_{2}$ appears. We are thus left with
\begin{equation}
\dot A=KA^{2}+B,
\label{dint23}
\end{equation}
where
\begin{equation}
K={\alpha_c^{3}\int_{0}^{\alpha_c}{\chi\over\xi^{2}}F{dF\over
d\xi}d\,\xi\over\int_{0}^{\alpha_c}{\chi} F \,d\xi}, \qquad B=\alpha_c
T_{c}^{2}\ {\int_{0}^{\alpha_c}\chi\xi^{2}\psi' e^{-\psi}
d\,\xi\over\int_{0}^{\alpha_c}{\chi} F\, d\xi}.
\label{dint25}
\end{equation}
For future convenience, we note that
\begin{equation}
F'(\xi)=\xi^{2}e^{-\psi}(2-\xi\psi').
\label{dint25F}
\end{equation}
By construction, the function $\chi$
satisfies the integral relation
\begin{equation}
\int_{0}^{\alpha_c} \biggl \lbrack
{\partial^{2}M_{2}\over\partial\xi^{2}}-{2\over\xi}{\partial
M_{2}\over\partial
\xi}+{1\over\xi^{2}}{\partial\over\partial\xi}(M_{2}\xi^{2}\psi')\biggr
\rbrack \chi(\xi)\,d\xi=0.
\label{dint26}
\end{equation}
Integrating by parts and using $M_{2}(0)=M_{2}'(0)=M_{2}(\alpha)=0$,
we obtain
\begin{equation}
\chi(\alpha_c)M_{2}'(\alpha_c,\tau)+\int_{0}^{\alpha_{c}} \biggl
\lbrace{d^{2}\chi\over\partial\xi^{2}}+\biggl ({2\over\xi}-\psi'\biggr
){d\chi\over d\xi}+{2\over\xi^{2}}(\xi\psi'-1)\chi\biggr \rbrace
M_{2}(\xi,\tau)\,d\xi=0.
\label{dint27}
\end{equation}
Thus, we impose that $\chi$ is a  solution of the differential equation
\begin{equation}
{d^{2}\chi\over\partial\xi^{2}}+\biggl ({2\over\xi}-\psi'\biggr
){d\chi\over d\xi}+{2\over\xi^{2}}(\xi\psi'-1)\chi=0,
\label{dint}
\end{equation}
with the boundary condition
\begin{equation}
\chi(0)=0.
\label{dint28}
\end{equation}
It turns out that the solution of this equation can be expressed
in terms of the Emden function as
\begin{equation}
\chi(\xi)={1\over\xi^{2}}F(\xi)e^{\psi}=\xi-\psi' e^{\psi},
\label{dint29}
\end{equation}
with $\chi'(0)=2/3$. The function $\chi(\xi)$ is represented in Fig.~3
along with the function $F(\xi)$. Using Eq. (\ref{dintconda}), we
readily check that $\chi(\alpha_{c})=0$. Therefore, the condition
Eq.~(\ref{dint26}) is satisfied. Accordingly, $A(\tau)$ is determined
by Eq.~(\ref{dint23}), where $K$ and $B$ have to be evaluated using
Eq.~(\ref{dint25}).

\begin{figure}[htbp]
\vskip1.5cm
\centerline{ \psfig{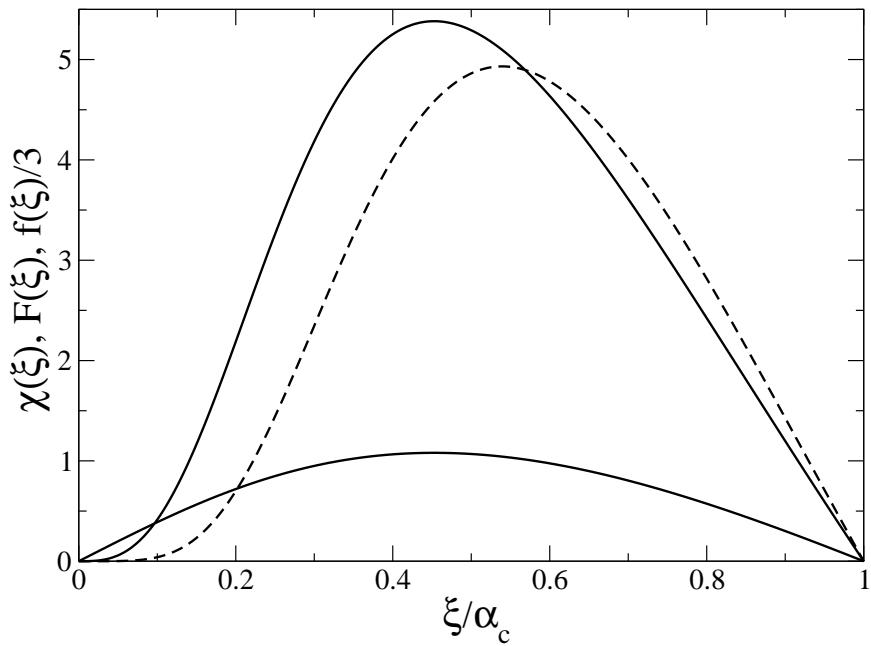}}
\caption{We plot $\chi(\xi)$ (lower curve), $F(\xi)$ (upper full
line) and the eigensolution $f(\xi)$ of Eq.~(\ref{eqmu}) (dashed
line ; $f(\xi)$ has been divided by a factor 3 for convenience).}
\label{chiFxi}
\end{figure}

\subsection{The blow up time}

The solution of Eq.~(\ref{dint23}) with $A(\tau)\rightarrow -\infty$ as
$\tau\rightarrow 0$ is
\begin{equation}
A(\tau)=\biggl ({B\over K}\biggr )^{1/2}{\rm tan}\biggl \lbrack \tau
(BK)^{1/2}-{\pi\over 2}\biggr\rbrack.
\label{dint30}
\end{equation}
Returning to the original time variable, we get
\begin{equation}
A(t)=\biggl ({B\over K}\biggr )^{1/2}{\rm tan}\biggl \lbrack
t(1-T/T_{c})^{1/2} (BK)^{1/2}-{\pi\over 2}\biggr\rbrack.
\label{dint31}
\end{equation}
Then, we find that the  blow up occurs for
\begin{equation}
t_{coll}=t_{*} (\eta-\eta_{c})^{-1/2},
\label{dint32}
\end{equation}
with
\begin{equation}
t_{*}=\pi\biggl ({\eta_{c}\over BK}\biggr )^{1/2}.
\label{dint33}
\end{equation}
We find numerically that
$K=62.56038...$ and $B=0.4716274...$. We thus get
\begin{equation}
t_{coll}=t_{*} (\eta-\eta_{c})^{-1/2},\qquad t_{*}=0.91767702...
\label{dint34}
\end{equation}
This asymptotic result is compared in Fig.~4 with the exact law
$t_{coll}(T)$ obtained by solving numerically the
Smoluchowski-Poisson system.

\vskip 1.2cm
\begin{figure}[htbp]
\centerline{ \psfig{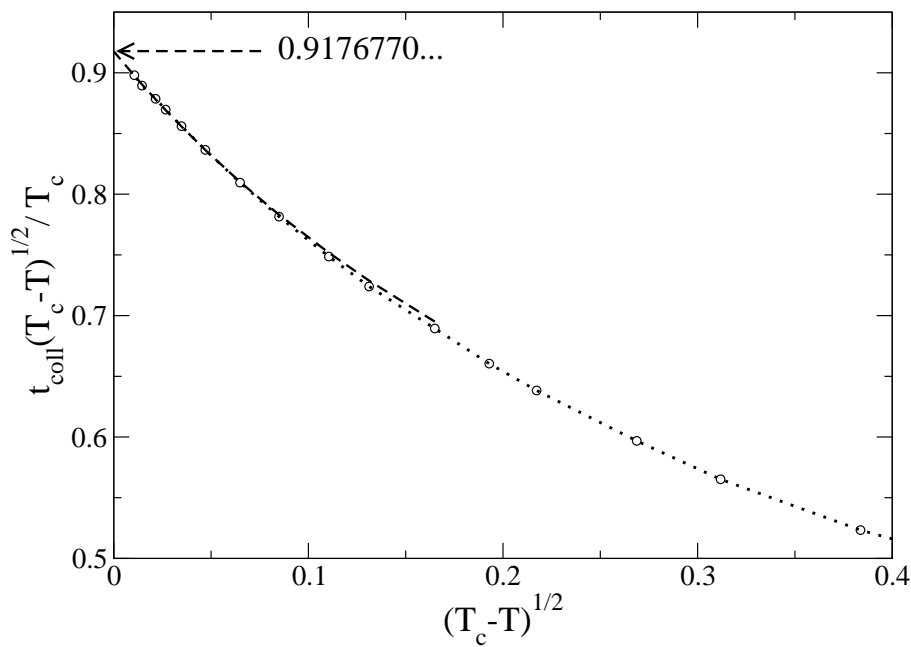}}
\caption{We plot $t_{coll}\times (T_c-T)^{1/2}/T_c$ computed
numerically as a function of  $(T_c-T)^{1/2}$, which should
converge to $t_*=0.91797702...$. A quadratic fit in the region
$(T_c-T)^{1/2}<0.05$ retrieves the first four digits of $t_*$.}
\label{figtcoll}
\end{figure}

\section{Relaxation time estimate for $T>T_c$}
\label{sec_relax}

We now assume $T>T_c$, so that an equilibrium state exists.  We look
for a time dependent solution for $M(r,t)$ which converges
exponentially to the equilibrium profile
\begin{equation}
M(r,t)=M_{T}(r)+F_T(r){\rm e}^{-t/\tau}.
\end{equation}
The purpose of this section is to compute the leading asymptotic form
of the relaxation time $\tau(T)\rightarrow +\infty$ as $T$ goes to
$T_c$.  After inserting this $ansatz$ into the equation of motion for
$M(r,t)$, we obtain
\begin{equation}
F_T''+\left(\phi'-\frac{2}{r}\right) F'_T+\alpha^2{\rm
e}^{-\phi}F_T=-(\tau T)^{-1}F_T, \label{eqprop}
\end{equation}
where
\begin{equation}
\phi(r)=\psi(\alpha r).
\end{equation}
We have used $r$ as a coordinate (instead of $\xi$) so as to remain
within a fixed interval of space $0\le r\le 1$ as we make the
perturbative expansion in $\alpha$ (see below). When $T\to T_c$, we
have $\alpha\to \alpha_c$, $\tau\to +\infty$, and $F_T(r)\to
F(\alpha_c r)$, where $F(\xi)$ is given by Eq.~(\ref{dint21}).  We now expand
Eq.~(\ref{eqprop}) in power of
$\varepsilon=\frac{\alpha_c-\alpha}{\alpha_c}$, by introducing
$f(\xi)$ and $\mu$ such that
\begin{equation}
F_T(r)=F(\alpha_c r)+\varepsilon f(\alpha_c r)+O(\varepsilon^2).
\end{equation}
and
\begin{equation}
(\tau\alpha_c^2T_c)^{-1}=\mu\varepsilon+O(\varepsilon^2).
\end{equation}
The relation between $T-T_c$ and $\varepsilon$ will
be given later.
Collecting terms proportional to $\varepsilon$, and setting
$\xi=\alpha_c x$, we obtain
\begin{equation}
f''+\left(\psi'-\frac{2}{\xi}\right)f'+f{\rm
e}^{-\psi}-\frac{2}{\xi^2}FF'=-\mu F,
\label{eqmu}
\end{equation}
subject to the boundary condition $f(0)=f(\alpha_c)=0$. This
eigenvalue problem selects a unique value for $\mu$, hence for
$\tau$. We were not able to solve this problem analytically.  However,
we remark that this equation is similar to Eq. (\ref{dint22}), without
the term leading to $B$. We can therefore use the same trick as in
Sec. \ref{sec_two}. After multiplying Eq.~(\ref{eqmu}) by the function
$\chi$ defined in the previous section, we end up with
\begin{equation}
\mu=\frac{\int_{0}^{\alpha_c}\frac{2\chi(\xi)}{\xi^2}F(\xi)F'(\xi)\,d\xi}
{\int_{0}^{\alpha_c} \chi(\xi)F(\xi)\,d\xi}=\frac{2K}{\alpha_c^3}=0.17202792...
\end{equation}
The direct numerical solution of Eq.~(\ref{eqmu})  leads of course
to the same value. The function $f(\xi)$ is plotted in Fig.~3.

We now find a simple relation between $T-T_c$ and $\varepsilon$ close
to $T_c$ \cite{aa}. Using the condition $\alpha\psi'(\alpha)=1/T$ and
the fact that $(\xi\psi')_{|_{\alpha_c}}'=0$, we obtain
\begin{equation}
\frac{1}{T_c}-\frac{1}{T}=\frac{(\alpha_c-\alpha)^2}{2}
(\xi\psi')_{|_{\alpha_c}}''+O((\alpha_c-\alpha)^3),
\end{equation}
which up to leading order in $\varepsilon$ can be written in the form
\begin{equation}
\varepsilon\sim\sqrt{\frac{2(T-T_c)}{T_c(1-2T_c)}}.
\end{equation}
Finally, we obtain the following diverging behavior for $\tau$:
\begin{equation}
\tau^{-1}\sim\mu\sqrt{\frac{2T_c(T-T_c)}{1-2T_c}}\alpha_c^2.
\end{equation}
Numerically, we get
\begin{equation}
\tau\sim c_T(T-T_c)^{-1/2}\sim c_\eta(\eta_c-\eta)^{-1/2}
\end{equation}
with
\begin{equation}
c_T=0.036563056...,\qquad c_\eta=0.09204937...
\end{equation}
Note that a heuristic argument leading to $\tau\sim (T-T_c)^{-1/2}$
was given in \cite{crs}, although the constant $c_T$ could not be
calculated.

\section{The decay of the density profile at $T=T_c$}

Strictly at $T=T_c$, $\tau=+\infty$, and one expects  a slow
convergence to the equilibrium profile $M_c(r)$.
Let us write
\begin{equation}
M(r,t)=M_c(r)-A(t)M_1(r)-B(t)M_2(r)+...
\label{devt}
\end{equation}
We implicitly assume that $B(t)\ll A(t)$ as $t\to +\infty$.  Moreover,
the presence of the minus signs in Eq.~(\ref{devt}) anticipates the
fact that starting from a flat density the central density
increases. In \cite{crs}, it was argued based on a heuristic argument
that $A(t)\sim t^{-1}$. It is the purpose of this section to justify
this statement as well as to give more quantitative estimates.

Let us postulate for the moment the natural
choice $B(t)\sim A(t)^2$ which will be justified hereafter.
Inserting again this $ansatz$ in the dynamical equation for
$M(r,t)$, and collecting terms of order $A(t)$ and $A(t)^2$, we
immediately find that the first equation leads to
\begin{equation}
M_1(r)=\sqrt{\frac{2T_c}{\alpha_c}}F(\alpha_c r),
\label{m1mu}
\end{equation}
where the constant has been set for later convenience, as multiplying $M_1$
and dividing $A$ by the same constant leaves the expansion for $M$
unchanged.  Setting $M_2(r)=g(\alpha_c r)$, we find that $g$ satisfies
\begin{equation}
g''+\left(\psi'-\frac{2}{\xi}\right)g'+g{\rm
e}^{-\psi}-\frac{2}{\xi^2}FF'=-\lambda F,
\label{eqlambda}
\end{equation}
with
\begin{equation}
\lambda=-\frac{\dot A}{A^2}\sqrt{\frac{2\alpha_c}{T_c}}\alpha_c^{-3},
\end{equation}
which must be a constant independent of time.  Thanks to the proper
choice of the constant in Eq.~(\ref{m1mu}), we find that
Eq.~(\ref{eqlambda}) and Eq.~(\ref{eqmu}) are identical. The only
physical solution with $g(0)=g(\alpha_c)=0$ corresponds to the
eigenvalue $\lambda=\mu$. Finally, we find
\begin{equation}
A(t)\sim t^{-1} \mu^{-1}\sqrt{\frac{2\alpha_c}{T_c}}\alpha_c^{-3}.
\label{atmu}
\end{equation}
Once the signs have been fixed in Eq.~(\ref{devt}), we indeed find that
the small time correction to the central density is necessarily
negative, as $M_1$, $M_2$ (or $g$) and $A$ are found to be positive. Therefore,
the equilibrium profile $M_{c}(r)$ is approached from below, which is natural since an excess of mass would provoke gravitational collapse.

Let us illustrate this result by calculating the correction to the
central density. Using Eq.~(\ref{m1mu}) and Eq.~(\ref{atmu}), and the
fact that $M_c(r)\sim\frac{4\pi}{3}\rho_0r^3$ and $F(\xi)\sim
\frac{2}{3}\xi^3$, we find that for large time the central density
converges to $\rho_0$ from below in a universal manner
\begin{equation}
\rho_0-\rho(r=0,t)=\frac{c_\rho}{t}+O(t^{-2}),
\end{equation}
with
\begin{equation}
c_\rho=(\pi\mu)^{-1}=1.8503385...
\end{equation}
This result is illustrated quantitatively in Fig.~5.

\begin{figure}[htbp]
\vskip1.5cm
\centerline{ \psfig{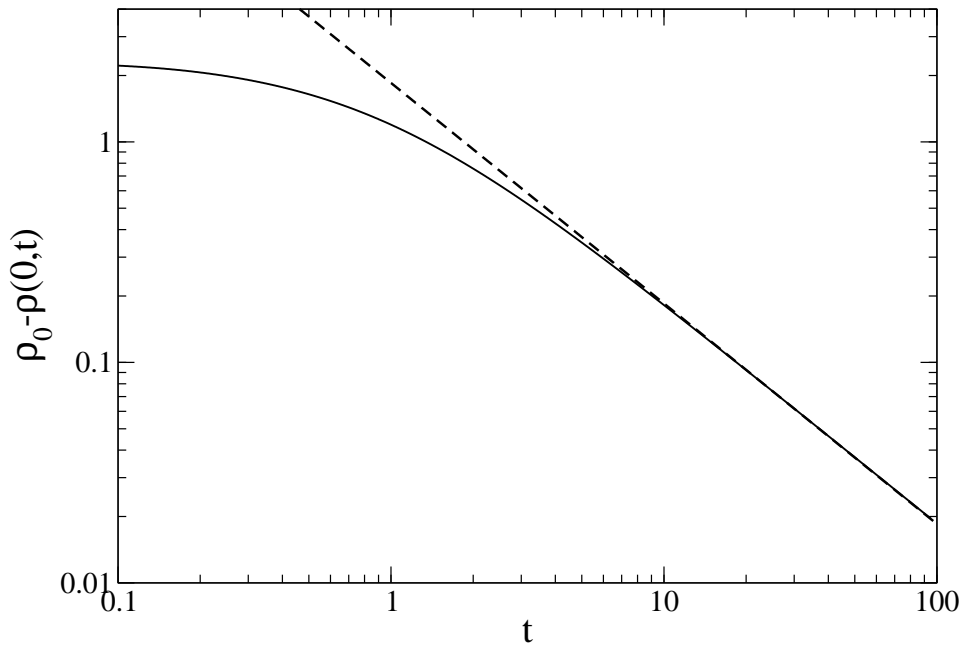}}
\caption{At $T=T_c$, and starting from a uniform mass density, we
have computed numerically $\rho_0-\rho(r=0,t)$, where $\rho_0$ is
the equilibrium central density. This is compared with the exact
and universal large time estimate $c_\rho/t$, with
$c_\rho=(\pi\mu)^{-1}=1.8503385...$ (dashed line).}
\label{rho0fig}
\end{figure}

Finally, we discuss the {\it a priori} unjustified choice
$B(t)\sim A^2(t)$. The case $B(t)\ll A^2(t)$ is clearly impossible
as it amounts to take $g=0$ in Eq.~(\ref{eqlambda}), leading to an
equation which is never satisfied by $F$, whatever the choice for
$A(t)$.  The opposite case $B(t)\gg A^2(t)$ leads to an equation
similar to Eq.~(\ref{eqlambda}) but where the term proportional to
$FF'$ is absent. For this equation, it is clear that if a solution
$g_0$ exists for a specific $\lambda_0$, then
$\lambda/\lambda_0g_0$ is also solution associated to $\lambda$.
This implies that the only solution satisfying the boundary
condition $g(0)=g(\alpha_c)=0$ is actually the null function
associated to $\lambda=0$, which is not permitted for $T>T_c$.
Hence $B(t)\sim A^2(t)$ is the only possibility leading to a
physically sensible solution.

\section{The pulsation period of isothermal spheres}

In Sec. \ref{sec_relax}, we have determined the relaxation time of self-gravitating Brownian particles described by the Smoluchowski-Poisson system close to the critical point $T_{c}$. Writing the perturbation as $\delta\rho\sim e^{\lambda t}$ where $\lambda=-1/\tau$, the eigenvalue equation for $\lambda$ can be written in dimensional form as \cite{crs}:
\begin{equation}
{d\over dr}\biggl ({1\over 4\pi \rho r^{2}}{dq\over dr}\biggr )+{Gq\over Tr^{2}}={\lambda\xi\over 4\pi \rho Tr^{2}}q,
\label{gs1}
\end{equation}
where $q(r)=\delta M(r)$. This equation is the same as
Eq. (\ref{eqprop}). Using the result of Sec. \ref{sec_relax}, and
returning to dimensional parameters, the inverse relaxation time is
given by
\begin{equation}
\lambda=\mp{1\over c_{\eta}}(\eta_{c}-\eta)^{1/2}{GM\over\xi R^{3}}.\label{gs2}
\end{equation}
The sign $-$ corresponds to $\alpha<\alpha_{c}$ and leads to exponentially damped pertubations (stable). The sign $+$ corresponds to $\alpha>\alpha_{c}$ and leads to exponentially growing pertubations (unstable).

We now consider the case of gaseous self-gravitating systems (stars)
described by the Euler-Jeans equations:
\begin{equation}
{\partial\rho\over\partial t}+\nabla(\rho {\bf v})=0,
\label{gs3}
\end{equation}
\begin{equation}
{\partial {\bf v}\over\partial t}+({\bf v}\nabla){\bf v}=-{1\over\rho}\nabla p-\nabla\Phi,
\label{gs4}
\end{equation}
\begin{equation}
\Delta\Phi=4\pi G\rho.
\label{gs5}
\end{equation}
We use an isothermal equation of state $p=\rho T$ and we assume that
the system is confined within a region of size $R$. These equations
provide the usual starting point for the analysis of the gravitational
stability of gaseous systems. The case of an infinite homogeneous
medium was first investigated by Jeans
\cite{jeans} in his classical study. The case of a finite
inhomogeneous medium was considered by Chavanis \cite{aa}. In that
case, the equation of pulsation can be written in the form
\begin{equation}
{d\over dr}\biggl ({1\over 4\pi \rho r^{2}}{dq\over dr}\biggr )+{Gq\over Tr^{2}}={\lambda^{2}\over 4\pi \rho Tr^{2}}q,
\label{gs6}
\end{equation}
where, as before, $\delta\rho\sim e^{\lambda t}$. Comparing with Eq. (\ref{gs1}), we see that the results of Sec. \ref{sec_relax} remain valid provided that $\lambda\xi$ is replaced by $\lambda^{2}$. Therefore, Eq. (\ref{gs2}) translates into 
\begin{equation}
\lambda^{2}=\mp{1\over c_{\eta}}(\eta_{c}-\eta)^{1/2}{GM\over R^{3}}.\label{gs7}
\end{equation}
For  $\alpha>\alpha_{c}$, the eigenvalue $\lambda=\pm \sqrt{\lambda^{2}}$ can be positive implying instability. For $\alpha<\alpha_{c}$, the eigenvalue $\lambda=\pm i\sqrt{-\lambda^{2}}$ is purely  imaginary so that the time evolution of the perturbation has an oscillatory nature. Close to the critical point, the pulsation period is given by
\begin{equation}
\omega={1\over \sqrt{c_{\eta}}}(\eta_{c}-\eta)^{1/4}\biggl ({GM\over R^{3}}\biggr )^{1/2}.\label{gs8}
\end{equation}

\section{Conclusion}

In this paper, we have obtained the asymptotic expressions of the
blow-up time and relaxation time of self-gravitating Brownian
particles and biological populations close to the critical point of
collapse in $d=3$. An excellent agreement is obtained by solving
numerically the Smoluchowski-Poisson system. This study confirms and
improves the results obtained in
\cite{crs} on the basis of heuristic arguments. A possible extension of
this work would be to consider the case of polytropic distributions
arising in case of anomalous diffusion \cite{anomalous}. More
generally, we could consider the generalized Smoluchowski equation (81)
proposed in \cite{gtpre} which is valid for an arbitrary equation of
state. This equation can provide (among other applications) a
generalized chemotactic model valid when the diffusion coefficient or
the drift term depends on the density. These extensions will be
considered in future works.

\section{Acknowledgements}

One of us (P.H.C) acknowledges interesting discussions with
N.I. Kavallaris and D.E. Tzanetis during the workshop on Nonlocal
elliptic and parabolic problems held in Bedlewo.

\end{document}